\documentclass[11pt, oneside]{article} 
\usepackage{afterpage}
\usepackage{epsfig}
\usepackage[a4paper]{geometry}
\usepackage{graphicx}	
\newcommand\ignore[1]{}			

\usepackage{amssymb}
\usepackage{amsmath}
\usepackage{amssymb}
\usepackage{amsthm}
\usepackage{graphicx}
\usepackage[urlcolor=blue,colorlinks=true,linkcolor  = black]{hyperref}
\usepackage{placeins}
\usepackage{tocloft}
\usepackage{slashed}
\usepackage{float}
\usepackage{fullpage}
\usepackage{braket}
\usepackage{url}

\usepackage{tabularx}

\usepackage{cite}
\usepackage{subcaption}
\usepackage{upgreek}
\usepackage{outlines}
\usepackage{braket}
\usepackage{appendix}
\def\0{{(0)}}
\def\1{{(1)}}
\usepackage{xspace}

\usepackage{mathrsfs}

\title{On the Origin and Fate of Our Universe}
\author{Cumrun Vafa\\ Harvard University}
\date{Talk presented at Lemaitre Conference 2024}

\begin{document}

\maketitle

\begin{abstract}
This brief review, intended for high energy and astrophysics researchers, explores the implications of recent theoretical advances in string theory and the Swampland program for understanding bounds on the structure of positive potentials allowed in quantum gravity. 
 This has a bearing on both inflationary models for the early universe as well as the fate of our universe.   The paper includes a review of the
dS conjecture as well as the TransPlanckian Censorship Conjecture (TCC) and its relation to the species scale.  We provide evidence for these principles as well as what they may lead to in terms of phenomenological predictions. 
\end{abstract}

\section{Introduction}
The discovery of dark energy and the accelerated expansion of the universe poses one of the greatest challenges to modern theoretical physics.  It signifies the existence of positive energy per unit volume of $\Lambda \sim 10^{-122}$ in Planck units.  Can this be reconciled with a UV complete description of quantum gravity, such as string theory?  More generally, what restrictions, if any, are there on small positive stable or flat potentials $V(\phi^i)$ as a function of scalar fields $\phi^i$?  A similar question, at much higher energies, also arises in the context of inflationary models of the early universe.   Clearly these are among some of the most important questions that a complete theory of quantum gravity needs to address.  One might naively expect that any positive scalar potential is allowed, but this is against what we have learned from string theory.  Indeed, the Swampland program (see \cite{Agmon:2022thq} for a recent review) seems to point towards having only a finite number of consistent theories of quantum gravity, and if so, clearly not all potentials are allowed.  The aim of this note is to explain what lessons we have learned from string theory in this regard and to discuss the restrictions on $V>0$ from the Swampland principles.  As far as the Swamplandish predictions for the fate of our universe, this review mainly focuses on the relative stability of the dark energy.  In a sense this review is complementary to the review in \cite{Vafa:2024fpx} which discusses Swamplandish predictions motivated by the smallness of the dark energy.

The organization of this paper is as follows.  In section 2, we review some of the basic theoretical background needed, and more specifically the distance conjecture (motivated from string dualities) and the species scale, related to black hole physics when there is a large degeneracy of light states. In section 3, we discuss the dS Conjecture and the TransPlanckian Censorship Conjecture (TCC) and use these and the ideas discussed in section 2, to place bounds on positive potentials.  In section 4, we examine the implications of these bounds for the origin of our universe and in particular inflation, as well as discuss Swampland motivated alternatives to inflation.  In section 5, we discuss the implications of these ideas for the fate of our universe.  We conclude this note in section 6.

\section{Theoretical Background}
\subsection{String dualities and the distance Conjecture}
String theory in supersymmetric backgrounds, typically has a number of massless fields, and more precisely with vanishing potential $V(\phi^i)=0$.  These fields constitute what is called the moduli space ${\cal M}$ of the theory. The kinetic term for these fields 
$$ {1\over 2}g_{ij}(\phi)\ \partial_\mu \phi^i \partial^\mu \phi^j$$
defines a natural metic $g_{ij}(\phi)$ on the moduli space ${\cal M}$.  Throughout this paper we will be working in Planck units. It has been known through all the known examples of quantum gravitational theories furnished by string theory that at large distances in field space, which for a canonically normalized field corresponds to $\phi\gg {\cal O}(1)$, light states emerge whose mass goes down exponentially with distance.  This is known as the distance conjecture \cite{Ooguri:2006in} and is expected to be generally true in all consistent theories of quantum gravity.  The light fields that emerge interact weakly with one another and lead to a {\it dual description} of the theory.
One finds through all known examples in string theory that there are only two possible weak coupling duals:  Either when some dimensions open up and decompactify the $d$ dimensional theory to a higher dimension $D$, or a tower of light strings appears.  That these are the only possibilities is known as the `Emergent String Conjecture'\cite{Lee:2019wij}.  One finds that
$$m\sim {\rm exp}(-\alpha \phi)\quad \phi\gg 1$$
(measured using the metric $g_{ij}(\phi)$) and in the first case 
$$\alpha \sim \frac{\sqrt{D-2}}{\sqrt{(D-d)(d-2)}}$$
and in the latter case
$$\alpha \sim \frac{1}{ \sqrt{d-2}}.$$
Note that both cases are consistent with a lower bound on the exponent $\alpha \geq \frac{1}{\sqrt {d-2}}$ \cite{Etheredge:2022opl}.  In fact more specifically the range for $\alpha$ is bounded:  $\frac{\sqrt{d-1}}{\sqrt{d-2}} \geq \alpha \geq \frac{1}{\sqrt {d-2}}$ where the upper bound comes from $D=d+1$. 
\subsection{Species Scale}
The appearance of a large number of light particles at large distances in field space implies that the black hole description should break down earlier at a lower energy scales (larger distance scales) than Planck scale, to avoid clash with the black hole entropy of a small black hole \cite{Dvali:2007hz,Dvali:2012uq}. Otherwise this would be in conflict with black hole entropy for a Planck scale black hole which is expected to account for all the light states.  In particular if there are $N$ light states, one expects the smallest black hole to be bigger (in Planck units) than
$l_{min}^{d-2} \gtrsim N$.  The corresponding energy scale $\Lambda_s=1/l_{min}$ is known as the species scale.  The fact that the black hole description should break down at smaller length scales is reflected by the fact that the gravitational action is expected to take the form \cite{vandeHeisteeg:2023ubh}
$$\int M_p^{d-2}({1\over 2}R+{R^2\over \Lambda_s^2}+{R^3\over \Lambda_s^4}+...)$$
where we have included some representative terms in the action. In string theory, given that asymptotically we expect a tower of light states which exponentially go down in mass one finds that $\Lambda_s(\phi)$  asymptotically also goes to zero exponentially 
$$\Lambda_s(\phi)\sim {\rm exp} (-\beta \phi)$$ 
and in the two limits of decompactificaiton limit $d\rightarrow D$ and weak coupling string limit they scale respectively as \cite{vandeHeisteeg:2023ubh}

$$\beta \sim \frac{\sqrt{D-d}}{ \sqrt{(D-2)(d-2)}}, \quad \beta \sim \frac{1}{\sqrt{d-2}}.$$
Species scale has been studied in diverse dimensions and a large body of evidence has emerged for the above statements \cite{vandeHeisteeg:2022btw,vandeHeisteeg:2023dlw,Calderon-Infante:2023uhz,Vafa:2024fti}.
Note that the emergent string conjecture therefore leads to a bounded range for $\beta$: $\frac{1}{\sqrt{(d-1)(d-2)}}\leq \beta \leq \frac{1}{\sqrt {d-2}}$ where the lower bound comes from $D=d+1$.

Note that the exponents appearing in the species scale seem somewhat related to the exponents we noted in the previous section for the mass of emerging tower of light states.  Indeed it has been found and explained based on the emergent string conjecture that quite generally we have \cite{Castellano:2023stg,Etheredge:2024tok} 
$$\frac{{\bf \nabla} m}{m}\cdot \frac{{\bf \nabla} \Lambda}{\Lambda} \rightarrow \frac{1}{d-2}$$
or in terms of the number of light species $N$,
$$\frac{{\bf \nabla} m}{m} \cdot \frac{{\bf \nabla}N}{N} \rightarrow -1.$$
The notion of the mass scale $m$ can be extended over the full moduli space and beyond asymptotic regime\footnote{In the two asymptotic limits $m$ corresponds to the Gregory-Laflamme transition temperature \cite{Gregory:1993vy} in one limit and the Horowitz-Polchinski transition temperature \cite{Horowitz:1997jc} in the other.} as signifying the scale where there is a phase transition in the structure of the black hole \cite{Bedroya:2024uva} in which case the above relation is no longer valid throughout the moduli and in particular can vanish at some points in the moduli.

\section{Bounds on V}
In this section we discuss what bounds arise for positive potential $V>0$ from the Swampland principles.  We first discuss the bound on the range of fields for flat regions of potential where $V\sim V_0$.  In the next section we discuss the dS conjecture and what implications it has for $V$ and finally we discuss the TransPlanckian Censorship conjecture, its motivation and its relation to the dS conjecture.
\subsection{Bounds on the Range of $V\sim V_0$}
The species scale places a bound on $V$ for the EFT to be a valid description.  In particular for $V$ to be reliably described by the EFT the curvature term it leads to should be smaller than $R\lesssim \Lambda_s^2$.  In other words for a valid EFT description we have
$$V\lesssim \Lambda_s^2(\phi)$$
Given that asymptotically $\Lambda_s(\phi)$ goes to zero exponentially at least with the rate $\beta \sim \frac{1}{\sqrt{(d-1)(d-2)}}$ corresponding to one extra dimension opening up $D=d+1$, and the fact that the ineterior of moduli space is of ${\cal O}(1)$ in Planck units, we find a bound for regions in field space for which $V$ is approximately constant $V\sim V_0$.  We  use $V_0\lesssim \Lambda_s^2$ to find that these regions are bounded by \cite{vandeHeisteeg:2023uxj}
$$\Delta \phi \lesssim \sqrt{(d-1)(d-2)}\ {\rm log}(\frac{1}{V_0}).$$
In other words to have a flat regions in potential over a long range requires having a small $V_0$.
\subsection{Bounds on $V>0$ and dS conjectures}

The string landscape with $V>0$ seems to have a very specific structure.  In particular in the weak coupling regimes (either weak coupling string, or large volume limits) which corresponds to large distance in field space potentials fall off exponentially:
$${\rm lim}_{\phi \gg 1}V(\phi)\sim {\rm exp} (-\gamma \phi)$$
where $\gamma \sim {\cal O}(1)$ in Planck units.   In all the string landscape examples one finds that asymptotically in field space $\gamma \geq {2\over {\sqrt{d-2}}}$ \cite{Rudelius:2022gbz,Andriot:2022xjh}.   Moreover, this implies that the slope of the potential is asymptotically too large to allow for inflationary expansion. That this may hold in all of the moduli space
is known as the dS conjecture \cite{Obied:2018sgi}.  A refined version of dS conjecture \cite{Garg:2018reu,Ooguri:2018wrx} states that for the entire field space, not only asymptotically,
$$\bigg|{V'\over V}\bigg|\geq {\cal O}(1)\quad or\ {\rm V \ is\ unstable:}  \ {V''\over V} <-{\cal O}(1).$$ Again these conditions go in the opposite direction to allowing inflation.  Moreover, if true, the dS conjecture would imply that there are no stable/metastable dS vacua.  The evidence for this conjecture, even though strong in the asymptotic regions, is not as strong in the interior regions of field space.

\subsection{TransPlanckian Censorship Conjecture}
TransPlanckian Censorship Conjecture (TCC) \cite{Bedroya:2019snp} is another, slightly different, version of the dS conjecture, which tries to relate the observed behavior of the potentials $V>0$ to a principle that one can imagine may have an explanation from first principles of a UV complete quantum gravitational theory.  The statement is that in a uniformly expanding universe with the scale factor $a(t)$, it should not be possible for a sub-Planckian region to become larger than the Hubble horizon $1/H(t)$ where $H=\dot{a}/a$.  In other words, in Planck units:
$${a(t_f)\over a(t_i)}\cdot 1 \leq {1\over H(t_f)}.$$
A motivation for this conjecture is that subPlanckian modes cannot be observable: if a subPlanckian mode crosses the horizon it classicalizes and in principle become observable if it enters back to our horizon.  What is surprising is that, if we assume this, it puts the following restrictions on the behaviour of $V$:  Asymptotically it leads exactly to the fact that if $V$ is exponentially falling off, the rate of fall off $\gamma \gtrsim \frac{2}{\sqrt{d-2}}$, exactly as it is observed in string theory setup!  This is a strong evidence for TCC, in that at least asymptotically it reproduces the observed string theory landscape bounds.  In fact we get yet another confirmation of TCC:  The bound on the region of flat potential coming from species scale considerations and its boundary behaviour, can also be derived from TCC, and one finds that the regions with almost flat potential $V_0$ is bounded by 
$$ \Delta \phi \lesssim \sqrt{(d-1)(d-2)}\ {\rm log}(\frac{1}{V_0})$$
To see this one uses the Friedmann equations in $d$ spacetime dimensions
$$\frac{(d-1)(d-2)}{2}H^2=\frac{1}{2}\dot\phi^2+V,$$
and the equation of motion for scalar field (which for simplicity we limit to one field):
$$\ddot\phi+(d-1)H\dot\phi+V'=0,$$
where $V'$ indicates the derivative of $V$ with respect to $\phi$.  Since $V>0$, from the Friedmann equation we deduce that
$$\frac{H}{|\dot\phi|}>\frac{1}{\sqrt{(d-1)(d-2)}}.$$
If we use the above lower bound for the integrand in the equation 
$$\int_{\phi_i}^{\phi_f}\frac{H}{\dot\phi}d\phi =\int_{t_i}^{t_f} Hdt<-\ln(H_f).$$

we find
$$\frac{|\phi_f-\phi_i|}{\sqrt{(d-1)(d-2)}}<-\ln(H_f),$$
which can be rearranged in the form
$$H_f<e^{-\frac{|\phi_f-\phi_i|}{\sqrt{(d-1)(d-2)}}}.$$
Due to the positivity of the kinetic term in the Friedmann equation this leads to
$$V<(d-1)(d-2)H^2/2\leq Ae^{-\frac{2}{\sqrt{(d-1)(d-2)}}|\phi-\phi_i|}$$
And combined with the fact that this is a bound on either side (where we assume potential falls off on both sides) we obtain the same result as we got using the species scale!

TCC leads to one more prediction.  Unlike the dS conjecture, TCC does not rule out a meta-stable dS vacuum.  However, its lifetime compared to dS scale would be rather short, so it is a matter of taste whether we should call this meta-stable or not. Indeed we find the upper bound on the age of such a universe $\tau$ is
$$\tau\lesssim \frac{1}{H}{\rm log}\frac{1}{H}$$
where $H^2\sim \Lambda$.  

In addition to the above motivations for TCC, there are further theoretical motivations for TCC, including from holographic principle; that in a quantum theory of gravity one should be able to describe the interior from asymptotic data \cite{Bedroya:2022tbh,Bedroya:2024zta}.
Even though TCC has similarities to dS conjecture, TCC seems better motivated and with more specific predictions.

\section{Implications for inflation}
 In order for inflation to provide a solution to
the structure formation problem of Standard Big Bang
cosmology, the current comoving Hubble radius must
originate inside the Hubble radius due to inflation.  This implies that
$$\frac{1}{H_i} \cdot e^{N_+}  \cdot \frac{T_R }{T_0} \gtrsim \frac{1}{H_0}$$
where $H_i$ is the Hubble scale during inflation, and $N_+$ is the number of e-foldings accrued during inflation after the CMB-scale modes exit the horizon $1/H$.
$T_0$ is the temperature of the CMB at the present time,
and $T_R$ is the corresponding temperature after reheating which we take to be $T_R\sim V^{1/4}\sim H_i^{1/2} $, where $V$ is the potential during inflation. $T_R/T_0$ in the above formula is the expansion factor since reheating till the present day.
Using the fact that now $H_0\sim T_0^2$, we arrive at
$$e^{N_+}\gtrsim \frac{V^{1/4}}{T_0}$$
Even without using TCC, and just using the species scale considerations 
 we could get a bound \cite{Scalisi:2018eaz,vandeHeisteeg:2023uxj} using the bound on the range of fields for regions of constant potential, as we have discussed.  Let us see what bound we get in this way. For simplicity,let us assume a single-field inflation. The range of inflaton movement during inflation is $\Delta \phi = \sqrt{2\epsilon} N_+$ where for slow-roll models $\epsilon \sim\frac{1}{2}(V'/V)^2$. Given that the region of flat potential is bounded in $d=4$ by
$$\Delta \phi \lesssim \sqrt{6}\ {\rm log}(1/V)$$
using $e^{N_+}\gtrsim \frac{V^{1/4}} {T_0}$ we find $V^{1/4+\sqrt{3/\epsilon}}<T_0$ and using the present value for $T_0\sim 10^{-32}$ leads for small $\epsilon$ to
$$V^{1/4}\lesssim 10^{-4.5 \sqrt{\epsilon}} $$
in Planck units which is not a strong bound given that $\epsilon$ is small. So the range of fields does not give a strong bound for inflation scale.

On the other hand enforcing TCC leads to a stronger restriction on inflationary models \cite{Bedroya:2019tba}.  Indeed, regardless of whether it is a single field or multi-field inflationary model, TCC demands that during inflation the expansion of the Planck scale cannot be more than the Hubble scale
$$\frac{a_f}{a_i}\cdot 1=e^{N_+}\cdot 1< \frac{1}{H} $$
which combining with the previous equation leads to
$$V^{1/4}\leq T_0^{1/3}\sim \Lambda^{1/12}\sim  10^{-10}\sim 10^9\  {\rm GeV}$$
where we used the present value of temperature $T_0\sim 10^{-32}$ in Planck units (which is similar to the energy slace for cosmological constant $\Lambda^{1/4}$). Using the observed power spectrum and equating it with curvature fluctuation
$$10^{-9}\sim {\cal P}\sim \frac{H^2}{\epsilon}$$
  Putting $H^2\lesssim 10^{-40}$
leads to 
$$\epsilon \lesssim 10^{-31},$$
which indeed looks highly fine tuned. In particular this leads to a small inflaton movement during inflation which would require highly fine tuned initial conditions. Moreover this small value for $\epsilon$ leads to tensor to scalar ratio $r= 16\epsilon \lesssim 10^{-30}$.  Given the rather fine tuned corner of inflationary model TCC drives us to, the naturalness in which inflation was supposed to resolve some of the puzzles of the early universe are put into question.  In particular assuming TCC renders inflation rather unnatural.
\subsection{A Swampland inspired alternative to inflation}
Given that at best the realization of inflation seems to be highly fine-tuned in the context of what we have learned from the Swampland program, it is natural to ask what alternatives are there for the early phase of our universe.
Motivated from string dualities one possibility was suggested in \cite{Brandenberger:1988aj} and more generally discussed in the context of a topological phase for the early universe in \cite{Agrawal:2020xek}.

The basic idea is that as we roll back the time the temperature increases and at some point the EFT should break down and a dual description should take over.  This is a prediction of string dualities:  There cannot be a single EFT description for any theory of quantum gravity valid in all regions of parameter space.  In particular at higher temperatures, in the Euclidean formulation, corresponds to the thermal circle of radius $1/T$ to become small, and for small radii we expect a dual formulation to take over.  Said differently, our phase of the cosmology would have arisen from a dual phase.  In such a picture the matter in our universe would get populated at the time of this conversion of ingredients from a dual phase to our phase.  This is the analog of the reheating phase in inflationary models.

This picture automatically solves many of the puzzles of the early universe.  For example the horizon problem, as to why apparently causally disconnected part of the universe have the same temperature, gets related to the fact that the conversion of the matter from the dual description to ours is insensitive to position defined in our frame, as the dual theory is not local relative to ours. Therefore this leads automatically to a conversion of dual matter density to a homogeneous one in our universe.  Our universe may potentially be described before such an epoch as a topological theory \cite{Agrawal:2020xek} related to a scale invariant topological gravity \cite{Witten:1988xi} and the emergence of our universe could emerge by breaking of this scale invariance.  This also leads, using unitarity, to a prediction of the red tilt in the power spectrum for deviations of scalar perturbations \cite{Agrawal:2020xek}, as has been experimentally observed.

\section{Implications for the fate of our Universe}
Observationally we know through the bounds on the dark energy variation that (see e.g. \cite{Agrawal:2018own})
$$\Big|\frac{V'}{V}\Big|\lesssim 0.5$$
If we assume dS conjecture, then $|V'/V|\not =0$.  However, we cannot be at the asymptotic limit of field space, because in that limit we have
$$\Big|\frac{V'}{V}\Big|\geq \sqrt{(d-2)}=\sqrt{2}$$
which would be in conflict with observations.  One naturally asks how close the slope can be to zero? 
Even though cannot be in an asymptotic regime of field space, if we also assume we are not at a meta-stable point (which TCC allows), it is natural to assume the slope is of ${\cal O}(1)$, such as one would expect in the refined dS conjecture in hilltop potentials \cite{Agrawal:2018rcg}. Indeed the recent preliminary results from DESI \cite{DESI:2024mwx} seem to go in the direction of finding non-vanishing slope for $V$.

On the other hand, if we assume TCC is correct, then meta-stable dS, where the slope vanishes, is also allowed.
Applying TCC to our universe leads, as already discussed, to a maximum lifetime for our universe:
 $$\tau^{max}\lesssim \frac{1}{H}{\rm log}\frac{1}{H}\sim \ 2 \ {\rm trillion}\ {\rm years}.$$
In other words our universe will not be lasting much beyond the Hubble scale, and so the meta-stable dS can hardly be called meta-stable, as it disappears in the time scale of dS.  In this way we see that the measurement of accelerated expansion could also be viewed as an indication that we are `near' the end of our universe!

\section{Conclusion}
Exploring the fate of our universe through the lens of string theory and the insights gained from the Swampland principles provides new leads to the expectation of instability of dark energy as well as leads to challenges for the inflationary paradigm. This framework deepens our understanding of cosmological aspects of our universe and can also potentially lead to observable effects involving the evolution of the dark energy component.

\subsubsection*{Acknowledgments}
We would like to thank Alek Bedroya, Damian van de Heisteeg, Georges Obied and Irene Valenzuela for helpful comments on this manuscript.

This work was supported in part by a grant from the Simons Foundation (602883, CV) and the DellaPietra Foundation.

\bibliographystyle{utphys}
\bibliography{references}

\end{document}